\documentclass[twocolumn,prl,aps]{revtex4}
\usepackage[dvips]{graphicx}
\begin{document}
\title{Polaronic behavior of undoped high-$T_c$ cuprates}
\begin{abstract}
We present angle-resolved photoemission spectroscopy (ARPES) data
on undoped La$_2$CuO$_4$, indicating polaronic coupling between bosons
and charge carriers.
Using a shell model, we calculate the electron-phonon coupling and find that it is strong enough to give polarons. We
develop an efficient method for calculating ARPES spectra in undoped
systems. Using the calculated couplings, we find the width of the phonon side band in good agreement with experiment.
We analyze reasons for the observed dependence of the width on the
binding energy.
\end{abstract}
\author{O.~R\"osch$^1$, O.~Gunnarsson$^1$,
X.~J.~Zhou$^{2,3}$, T.~Yoshida$^4$, T.~Sasagawa$^5$, A.~Fujimori$^4$,
Z.~Hussain$^3$, Z.-X.~Shen$^2$ and S.~Uchida$^4$
}
\affiliation{
$^1$Max-Planck-Institut f\"ur Festk\"orperforschung,
D-70506 Stuttgart, Germany
\\
$^2$Department of Applied Physics and Stanford Synchrotron Radiation Laboratory,
Stanford University, Stanford, California 94305, USA
\\
$^3$Advanced Light Source, Lawrence Berkeley National Lab, Berkeley,
California 94720, USA
\\
$^4$Department of Physics and Department of Complexity Science
and Engineering, University of Tokyo, Tokyo, 113-0033, Japan\\
$^5$Department of Advanced Materials Science, University of Tokyo, Japan
}

\maketitle

An important issue in the discussion of high-$T_c$ cuprates is the
relative importance of couplings to phonons and spin fluctuations.
Recent angle-resolved photoemission spectroscopy (ARPES) work for
undoped or weakly doped Ca$_{2{\textrm-}x}$Na$_x$CuO$_2$Cl$_2$ shows polaronic
behavior \cite{Na}. A broad ${\bf k}$-dependent boson side band
was observed, while a quasi-particle, if present, was too weak to be seen. 
This raises important questions about the nature of the bosons coupling
to the electrons. Here we present experimental data for La$_{2{\textrm-}x}$Sr$_x$CuO$_4$,
showing similar boson side bands. Using a shell model, we calculate 
the electron-phonon coupling for undoped La$_2$CuO$_4$. We find that
the coupling is sufficiently strong to give polaronic behavior. We
develop a method for calculating spectra of a $t$-$J$ model with many 
phonon modes and at finite temperature. The calculated couplings
give side band widths in good agreement with experiment. 
The dependence of the width on the binding energy is analyzed. This 
work shows that the electron-phonon coupling plays an important 
role for properties of undoped cuprates.

\begin{figure}
\includegraphics[width=8cm]{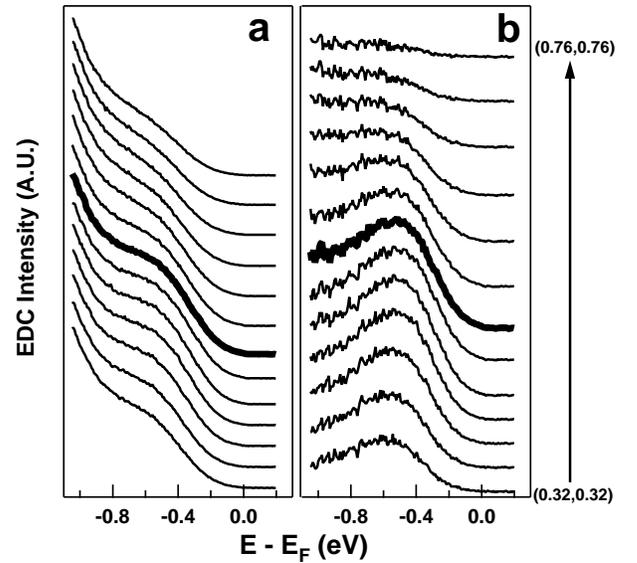}
\caption{\label{fig0}(a) Photoemission spectra of La$_2$CuO$_4$
along the (0,0)-($\pi$,$\pi$) nodal direction in the first
Brillouin zone. The corresponding momentum runs from (0.32,0.32)$\pi$ to
(0.76,0.76)$\pi$,  as indicated by the arrow. (b) To highlight
the momentum dependence, a "background" given by a spectrum near ($\pi$,$\pi$)
was subtracted from the spectra. The bold
curves correspond to a momentum near ($\pi$/2,$\pi$/2).}
\end{figure}

The photoemission measurements were carried out on beamline 10.0.1
at the Advanced Light Source. The photon energy was 55 eV and the
energy resolution was $\approx$~18~meV.  The La$_2$CuO$_4$ single
crystals were grown by the traveling solvent floating zone method. The
samples were cleaved {\it in situ} in vacuum with a base pressure
better than 4$\times$10$^{-11}$ Torr and measured at a temperature
of $\approx$~20~K. The data are shown in Fig.~\ref{fig0}.
There are broad boson side bands at about 0.5~eV binding energy,
similar to the results for Ca$_{2{\textrm-}x}$Na$_x$CuO$_2$Cl$_2$ \cite{Na}.

To see if these boson side bands can be explained by the coupling to phonons,
we calculate the electron-phonon interaction.
We use a shell model \cite{shell} that successfully describes phonon dispersions in several cuprates to obtain the phonon modes in La$_2$CuO$_4$. Within this model the change in the electrostatic potential induced by each mode is calculated to first order in the atomic displacements. This gives rise to a linear coupling between
doped charges and the phonons.
We use the $t$-$J$ model to describe the electronic degrees of freedom 
in the CuO$_2$ planes. Doped holes form Zhang-Rice singlets with the spins 
of Cu $d$ holes with their additional charge distributed symmetrically on the four
neighboring O $p$ orbitals.  This charge distribution couples to the modulation 
of the electrostatic potential  caused by the phonons.                               
Since the undoped system is an insulator, screening in the shell model is only provided by the displacement of the shells against the atomic cores.
In addition, there is coupling from the modulation of the $p$-$d$ hopping $t_{pd}$ and the charge-transfer energy in the three-band model from which the $t$-$J$ model is derived \cite{tJ}. We include the dominant on-site terms.
The modulation of the charge-transfer energy is calculated from the modulation of $p$ and $d$ level energies within the shell model  instead of assuming a nearest-neighbor Coulomb repulsion $U_{pd}$ as in Ref.~\cite{tJ}.
The shell model calculations are done on a $(30)^3$ mesh in $\bf q$ space using the high-temperature tetragonal structure from Ref.~\cite{shell} but similar results
are obtained for the low-temperature orthorhombic structure.
The results depend on the eigenvectors of the shell model. Since the shell model \cite{shell} describes neutron scattering intensities well, the eigenvectors are believed to be accurate \cite{pintsch}.

The electron-phonon interaction is then given by
\begin{equation}\label{Hep}
H_{ep}=\frac{1}{\sqrt{N}}
\sum_{{\bf q}\nu i}
g_{{\bf q}\nu}
(1-n_i)
\sqrt{2\omega_{{\bf q}\nu}}
Q_{{\bf q}\nu}
e^{i{\bf q\cdot R}_i}.
\end{equation}
This is an on-site coupling to empty sites representing singlets in the $t$-$J$ model. It is linear in the generalized phonon coordinates $Q_{{\bf q}\nu}$ and the strength is given by the coupling constants $g_{{\bf q}\nu}$. A phonon mode with eigenfrequency $\omega_{{\bf q}\nu}$
is labeled by its wave vector $\bf q$ and branch index $\nu$.
$n_i$ measures the occupancy at site ${\bf R}_i$. The number of sites is $N$.

Introducing the dimensionless coupling constant $\lambda\equiv2
\sum_{{\bf q}\nu}|g_{{\bf q}\nu}|^2/(8t\omega_{{\bf q}\nu}N)$, 
we obtain $\lambda=1.2$ \cite{mesh} for $t=0.4$~eV. The criterion 
for polaronic behavior in the Holstein-$t$-$J$ model is 
$\lambda>\lambda_c=0.4$ \cite{MN}. If the next-nearest neighbor 
hopping integral $t'$ is  taken into account, $\lambda_c$ should increase. The coupling 
derived for undoped La$_2$CuO$_4$ should, nevertheless, be strong 
enough to cause polaronic behavior, as observed experimentally.

\begin{figure}
\includegraphics[width=8cm]{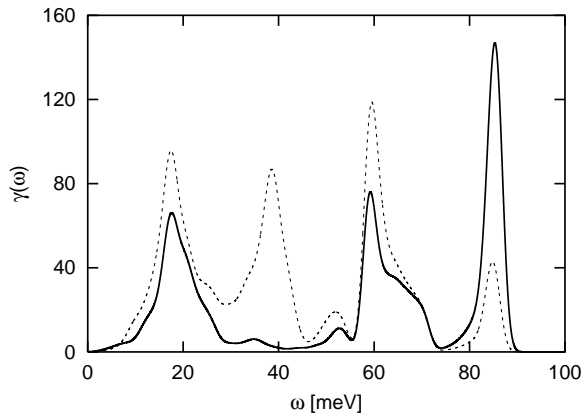}
\caption{\label{fig4}Spectral distribution of coupling to singlets (full line) and to holes in non-bonding O $p$ orbitals (dashed line).
A Gaussian broadening (FWHM=3 meV) was used.}
\end{figure}
Fig.~\ref{fig4} shows the spectral distribution
$\gamma(\omega)\!\equiv\!\sum_{{\bf q}\nu}|g_{{\bf q}\nu}|^2/(\omega_{{\bf q}\nu}N)\delta(\omega-\omega_{{\bf q}\nu})$ of our coupling (full line).
A peak around 20 meV can be attributed mainly to modes involving La whereas the spectral weight at 60-70~meV
is mainly due to vibrations of the apical O.
A peak around 85~meV is caused by the
planar O \mbox{(half-)breathing} mode coupling
predominantly via the $t_{pd}$ modulation.
This energy dependence of the coupling agrees qualitatively
with the observation of fine structures in the electron self-energy
of La$_{2{\textrm-}x}$Sr$_x$CuO$_4$ \cite{multi}, except
that the calculation does not give appreciable coupling around 40 meV.
This could be due to surface effects or distortions due to doping \cite{Bianconi,Takagi,Hackl}.
Such effects would be missing in our calculation, which is performed for an ideal La$_2$CuO$_4$ structure.
For comparison, the dashed line in Fig.~\ref{fig4} shows the spectral distribution of the coupling to holes in individual non-bonding
O $p$ orbitals.

To discuss the $\bf q$ dependence of the coupling
we sum $|g_{{\bf q}\nu}|^2$ over all modes and the phonon momentum $q_z$ perpendicular to the CuO$_2$ planes.
The coupling increases with decreasing
$|(q_x,q_y)|$ although some individual modes show different behavior, e.g. the coupling to the O \mbox{(half-)breathing} mode peaks around $(\pi,\pi)$.

Based on Ref.~\cite{EPJB} we have developed an efficient method for 
calculating ARPES spectra in undoped systems at a finite temperature $T$.
Using an adiabatic approximation, the spectral function for creating 
a hole with momentum ${\bf k}$ in an undoped system is written as
\begin{equation}\label{approx}
A_{\bf k}(\omega,T)
=
\int\!d{\bf Q}
\ |\phi_0({\bf Q},T)|^2
\rho_{\bf k}({\bf Q},\omega).
\end{equation}
The integration is over all phonon coordinates $Q_{{\bf q}\nu}$ 
summarized in the vector $\bf Q$.  $\rho_{\bf k}({\bf Q},\omega)$ 
is the spectrum calculated for the electronic system in a distorted 
lattice (characterized by $\bf Q$). No dynamic phonons are included 
in the calculation of $\rho_{\bf k}({\bf Q},\omega)$, since the 
distortions $\bf Q$ are treated as $c$-numbers and only give rise to 
static on-site energies in the Hamiltonian~(\ref{Hep}). 
\begin{equation}\label{wave}
|\phi_0({\bf Q},T)|^2=
\prod_{{\bf q}\nu} \sqrt{\frac{\omega_{{\bf q}\nu}}{\pi}}
\ \exp\left(\frac{-\omega_{{\bf q}\nu}Q_{{\bf q}\nu}^2}
{2n_{{\bf q}\nu}(T)+1}\right)
\end{equation}
is the squared ground-state wavefunction for non-interacting phonons at $T=0$ (or the corresponding ensemble average for finite $T$,
\mbox{$n_{{\bf q}\nu}(T)=(\exp(\omega_{{\bf q}\nu}/(k_BT))-1)^{-1}$} is the Bose-Einstein distribution).
Our approximation neglects the kinetic energy of the phonons in the final states and assumes that $k_BT$ is small compared with the electronic energy scale. We use the fact that there is no electron-phonon coupling for the initial state in the undoped $t$-$J$ model.

In practice, Eq.~(\ref{approx}) can be evaluated efficiently by Monte Carlo sampling over the phonon coordinates using $|\phi_0({\bf Q},T)|^2$ as the weighting function. Each sampled $\rho_{\bf k}({\bf Q},\omega)$ requires only the calculation of a purely electronic problem. There is no blow-up of the Hilbert space through the inclusion of dynamic phonons that usually prevents calculations for strong coupling.

To obtain ARPES spectra for La$_2$CuO$_4$ we describe the electronic system by the two-dimensional $t$-$J$ model choosing $t=0.4$ eV and $J=0.3t$. $\rho_{\bf k}({\bf Q},\omega)$ is calculated using exact diagonalization on a 4$\times$4 cluster with periodic boundary conditions.
To check our method we have first assumed a simple Holstein-type of coupling to one dispersionless mode as in Ref.~\cite{MN} and found good agreements with their results.
But our method allows us to include all 21 modes
with the $\bf q$-dependent coupling given by Eq.~(\ref{Hep}) \cite{eff}.

\begin{figure}
\includegraphics[width=8cm]{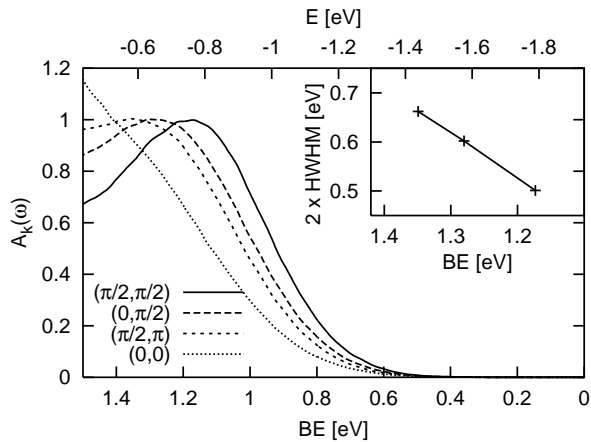}
\caption{\label{fig1}ARPES spectra for the undoped system at $T=0$ 
for different $\bf k$ normalized to the height of the phonon side band. 
The lower abscissa shows binding energies (BE) and the upper abscissa 
the energies of the final states corresponding to the spectral features.
The inset shows the dependence of the width of the phonon side band 
on its binding energy.  The width of the $(0,0)$ spectrum is poorly 
defined and not shown.
}
\end{figure}

We have calculated $A_{\bf k}(\omega,T=0)$ for
different values of $\bf k$ as shown in the main part of
Fig.~\ref{fig1}.
For each spectrum 50000 samples were used. The poles in the sampled spectra were broadened by Gaussians with a FWHM of 33~meV.
The spectra show a broad phonon side band
which disperses like the quasi-particle in the $t$-$J$ model without phonons, as was found in Ref.~\cite{MN}. As shown in Ref.~\cite{EPJB} this can be understood easily within our adiabatic approximation. The true quasi-particle peak is almost completely suppressed in weight and dispersion and determines the
energy zero on the lower abscissa in Fig.~\ref{fig1}.
The half-width (HWHM) of the side band is determined on the low binding energy side.
We obtain a width of $2\times$HWHM=0.5~eV for ${\bf k}=(\pi/2,\pi/2)$,
in good agreement with the corresponding experimental result 0.48~eV
(cf. bold curves in Fig.~\ref{fig0}).

\begin{figure}
\includegraphics[width=8cm]{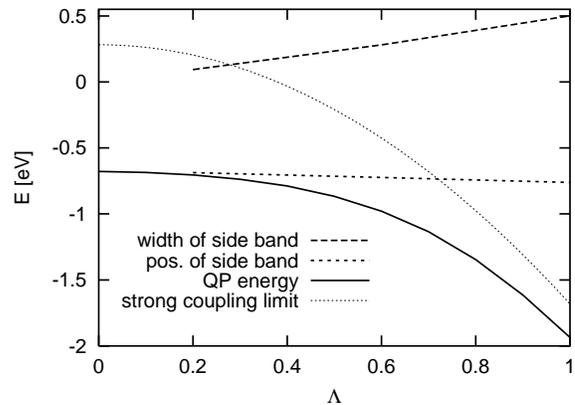}
\caption{\label{fig2}Width and position of the phonon side band and the position of the quasi-particle peak as a function of the relative coupling strength $\Lambda$ for ${\bf k}=(\pi/2,\pi/2)$.}
\end{figure}

Width and position of the phonon side band
as well as the position of the quasi-particle peak
are shown in Fig.~\ref{fig2}
as a function of the relative coupling strength $\Lambda$
(substituting $g_{{\bf q}\nu}\to\Lambda g_{{\bf q}\nu}$
in Eq.~(\ref{Hep})) for ${\bf k}=(\pi/2,\pi/2)$ at $T=0$.              
For small $\Lambda$, the position of the quasi-particle peak \cite{note}
starts out at the energy obtained in the $t$-$J$ model without phonons and approaches for stronger couplings asymptotically the curve given by
$-\sum_{{\bf q}\nu}\Lambda^2|g_{{\bf q}\nu}|^2/\omega_{{\bf q}\nu}+const.$
A fully localized hole obtains this energy gain from interaction with 
phonons (the additional constant depends on the definition of the 
energy offset). The position of the phonon side band shows only a weak 
linear dependence on $\Lambda$. The energy difference to the position of 
the true quasi-particle peak determines the binding energy. The value of 
almost 1.2~eV at full coupling is larger than the experimental value of 
about 0.5~eV (cf. Fig.~\ref{fig0}).
Since the binding energy depends strongly on doping,
a possible explanation for this discrepancy could be that the experimental samples
were actually lightly doped.
Our shell model calculation could also overestimate the coupling strength.
As seen in Fig.~\ref{fig2}, the binding energy of the side band has a much stronger
dependence on $\Lambda$ than the width. For $\Lambda=0.8$, e.g.,
one would find a width of about 0.4~eV and a binding energy of about 0.6~eV, both
quantities being in reasonable agreement with experimental values.
As $\lambda\propto\Lambda^2$, this would reduce the dimensionless coupling constant to
$\lambda=0.75$. Such a reduction could be due to a slight underestimate of the screening in the shell model or an overestimate of the coupling to breathing phonons.

The inset in Fig.~\ref{fig1} shows that the width of the phonon side band
increases roughly linearly with its binding energy.
This trend has also been found experimentally for weakly doped
Ca$_{2{\textrm-}x}$Na$_x$CuO$_2$Cl$_2$ \cite{Na}.
\begin{figure}
\includegraphics[width=6cm]{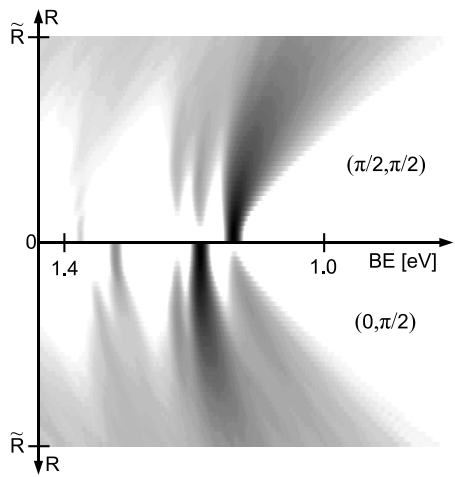}
\caption{\label{fig3} Contributions to the spectra with
${\bf k}=(\pi/2,\pi,2)$ and ${\bf k}=(0,\pi,2)$ at $T=0$ for varying $R$. Higher spectral weight is represented by darker shading.
The abscissa shows binding energies (BE). Coupling to ${\bf q}=(0,0)$ modes is not included.}
\end{figure}
To understand this, we introduce rescaled phonon coordinates
$R_{{\bf q}\nu}=\sqrt{\omega_{{\bf q}\nu}/(2n_{{\bf q}\nu}(T)+1)}Q_{{\bf q}\nu}$.
The weighting function in Eq.~(\ref{wave}) then becomes invariant under rotations in ${\bf R}$ space, and Eq.~(\ref{approx}) can be expressed as a sampling over directions for fixed $R=|{\bf R}|$ followed by an integration over $R$. If the $\bf R$ space has dimension $d$, the $R$-integral contains a factor $R^{d-1}e^{-R^2}$,
which peaks strongly at $\tilde R\equiv\sqrt{(d-1)/2}$. Fig.~\ref{fig3} shows 
how the contributions to the spectrum evolve with increasing $R$ up to $\tilde R$.
Poles in the sampled spectra were broadened by Gaussians with a FWHM of 11 meV.
In the figure, the coupling to the {\bf q}=(0,0) modes is not included, since it contributes the same broadening for all ${\bf k}$. We focus on peaks to the right, at the lowest binding energies. For $R=0$,
which corresponds to zero electron-phonon coupling,
the spectrum for ${\bf k}=(\pi/2,\pi/2)$
has a peak at smaller binding energy than for ${\bf k}=(0,\pi/2)$.
As $R$ increases and the electron-phonon coupling is switched on, both peaks are broadened. This is due to the repulsion from higher states, which increases with $R$, but is different for different directions of ${\bf R}$. 
For the peak with the lowest binding energy this broadening
is largest, since all the other states repel the corresponding state in the same direction. Fig.~\ref{fig1} shows, however, the opposite trend
for the width of the phonon side band in the final spectra, due to the following opposing effect which dominates. For $R\neq0$ spectral weight also appears at energies with no peaks in the $R=0$ spectra.
For $R\ne 0$, the system is distorted and the electronic momentum $\bf k$ is not conserved. For instance, in case of ${\bf k}=(0,\pi/2)$, a peak of increasing width and weight appears at the energy of the main peak in the ${\bf k}=(\pi/2,\pi/2)$ spectrum. At $R=\tilde R$, these side bands have more or less merged with the main peak and effectively add to its width. For ${\bf k}=(\pi/2,\pi/2)$ there is no such extra contribution for lower binding energies whereas for ${\bf k}=(0,\pi/2)$ the width is increased by one side band.
There are more and more side bands on the low binding energy side of the main peak as the binding energy of the main peak increases, which increases the width of the resulting broad peak on the low binding energy side.

Finally, we have studied the temperature dependence of the width of the phonon 
side band. It is relatively weak. For ${\bf k}=(\pi/2,\pi/2)$ it increases 
from 500~meV at $T=0$ to 565~meV at $T=200$~K and 750~meV at $T=400$~K. The 
main contribution to the $T$ dependence is due to the modes with an energy 
of about 20~meV.

To summarize, we have shown that the polaronic features in ARPES spectra 
of La$_2$CuO$_4$ and other undoped cuprates can be attributed to strong
coupling between the photo hole and phonons.
This electron-phonon coupling is due to the modulation of the electrostatic potential and the singlet energy. We have introduced an efficient method for obtaining ARPES spectra in undoped systems and shown that with the derived coupling there is good agreement with experimental results. The dependence of the width of the broad phonon side band on binding energy and temperature has also been discussed.
Our results show the importance of electron-phonon coupling for the physics of undoped cuprates.

We thank T. P. Devereaux and P. Prelov$\breve s$ek for useful discussions.


\begin{thebibliography}{99}
\bibitem{Na}
K. M. Shen, F. Ronning, D. H. Lu,
W. S. Lee, N. J. C. Ingle, W. Meevasana,
F. Baumberger, A. Damascelli, N. P. Armitage,
L. L. Miller, Y. Kohsaka, M. Azuma,
M. Takano, H. Takagi, and Z.-X. Shen,
Phys. Rev. Lett. {\bf 93}, 267002 (2004).

\bibitem{shell}
S. L. Chaplot, W. Reichardt, L. Pintschovius,
and N. Pyka,
Phys. Rev. B {\bf 52}, 7230 (1995).

\bibitem{tJ}
O. R\"osch and O. Gunnarsson,
Phys. Rev. Lett. {\bf 92}, 146403 (2004).

\bibitem{pintsch}
L. Pintschovius and W. Reichardt,
in {\it Neutron Scattering in Layered Copper-Oxide Superconductors},
edited by A. Furrer, Physics and Chemistry of Materials with
Low Dimensional Structures, Vol. 20 (Kluwer Academic, Dordrecht, 1998),
p. 165.

\bibitem{MN}
A. S. Mishchenko and N. Nagaosa,
Phys. Rev. Lett. {\bf 93}, 036402 (2004).

\bibitem{mesh}
The finite number of mesh points leads to an underestimate
of $\lambda$ by about $3\%$ for a $(30)^3$ mesh.

\bibitem{multi}
X. J. Zhou, J. Shi, T. Yoshida, T. Cuk, W. L. Yang,
V. Brouet, J. Nakamura, N. Mannella, S. Komiya, Y. Ando, F. Zhou, W. X. Ti,
J. W. Xiong, Z. X. Zhao, T. Sasagawa, T. Kakeshita, H. Eisaki, S. Uchida,
A. Fujimori, Z. Zhang, E. W. Plummer, R. B. Laughlin, Z. Hussain,
and Z.-X. Shen,
cond-mat/0405130.

\bibitem{Bianconi}
A. Bianconi, N. L. Saini, A. Lanzara, M. Missori, T. Rossetti,
H. Oyanagi, H. Yamaguchi, K. Oka, and T. Ito,
Phys. Rev. Lett. {\bf 76}, 3412 (1996).

\bibitem{Takagi}
E. S. Bozin, G. H. Kwei, H. Takagi, and S. J. L. Billinge,
Phys. Rev. Lett. {\bf 84}, 5856 (2000).

\bibitem{Hackl}
L. Tassini, F. Venturini, Q.-M. Zhang, R. Hackl, N. Kikugawa,
and T. Fujita,
cond-mat/0406169.

\bibitem{EPJB}
O. R\"osch and O. Gunnarsson,
Eur. Phys. J. B {\bf 43}, 11 (2005);
K. Sch\"onhammer and O. Gunnarsson,
Phys. Rev. B {\bf 30}, 3141 (1984).

\bibitem{eff}
The $q_z$ dependence is unimportant as the electronic model is purely two-dimensional. For given $q_x,q_y,\nu$ we therefore use a single mode
with effective coupling $g_{\rm eff}=\sqrt{\sum_{q_z}|g_{q_z}|^2}$ and
frequency $\omega_{\rm eff}=g_{\rm eff}^2/\sum_{q_z}(|g_{q_z}|^2/\omega_{q_z})$


\bibitem{note}
The position of the true quasi-particle peak is obtained from a simulated annealing search over the adiabatic energy surface.

\end{thebibliography}
\end{document}